\documentclass[twocolumn,prd,superscriptaddress,nofootinbib]{revtex4-1}
\usepackage{amsfonts,amsmath,amssymb,mathrsfs}
\usepackage{hyperref}
\usepackage{color}
\usepackage{graphicx}  
\usepackage{dcolumn}   
\usepackage{bm}        
\usepackage[table]{xcolor}
\usepackage[english]{babel}
\usepackage{comment}
\def\a{\alpha}

\def\r{\rho}
\def\s{\sigma}
\def\t{\tau}
\def\m{\mu}
\def\n{\nu}
\def\k{\kappa}
\def\th{\theta}
\def\g{\gamma}\def\G{\Gamma}
\def\L{t}\def\l{V}
\def\D{\Delta}
\def\la{\langle}
\def\ra{\rangle}
\def\o{\omega}\def\O{\Omega}
\def\d{\delta}
\def\p{\partial}

\def\oxthree{{\cal O}(x^3) }

\def\half{\textstyle{\frac{1}{2}}}

\def\bdoc{\begin{document}}
\def\edoc{\end{document}}
\def\bea{\begin{equation}}
\def\eea{\end{equation}}

\def\beq{\begin{eqnarray}}
\def\eeq{\end{eqnarray}}
\def\be{\begin{eqnarray}}
\def\ee{\end{eqnarray}}
\def\ben{\begin{enumerate}}
\def\een{\end{enumerate}}
\def\la{\langle}
\def\ra{\rangle}
\def\a{\alpha}
\def\g{\gamma}\def\G{\Gamma}
\def\d{\delta}\def\D{\Delta}
\def\e{\epsilon}
\def\z{\zeta}

\def\th{\theta}
\def\k{\kappa}
\def\l{t}
\def\m{\mu}
\def\n{\nu}
\def\o{\omega}
\def\p{\pi}
\def\r{\rho}
\def\s{\sigma}
\def\t{\tau}
\def\L{{\cal L}}
\def\S{\Sigma }
\def\gsim{\; \raisebox{-.8ex}{$\stackrel{\textstyle >}{\sim}$}\;}
\def\lsim{\; \raisebox{-.8ex}{$\stackrel{\textstyle <}{\sim}$}\;}
\def\gtrsim{\gsim}
\def\lessim{\lsim}
\def\loc{{\rm local}}
\def\vm{v_{\rm max}}
\def\bh{\bar{h}}
\def\del{\partial}
\def\nab{\nabla}
\def\half{{\textstyle{\frac{1}{2}}}}
\def\fourth{{\textstyle{\frac{1}{4}}}}

\def\bD{{\bf D}}
\def\bE{{\bf E}}
\def\bF{{\bf F}}
\def\bB{{\bf B}}
\def\bP{{\bf P}}
\def\bV{{\bf v}}
\def\bv{{\bf v}}
\def\bx{{\bf x}}
\def\by{{\bf y}}
\def\bz{{\bf z}}
\def\ba{{\bf a}}
\def\bd{{\bf d}}
\def\bs{{\bf s}}
\def\bn{{\bf n}}
\def\bp{{\bf p}}

\def\O{\Omega}

\def\br{{\bf r}}
\def\bnab{{\bf \nab}}

\def\tE{\tilde{E}}
\def\tL{\tilde{L}}
\def\Horava{Ho\v{r}ava }

\def\oxtwo{\mathscr{O}\left(x^2\right)}
\def\oxthree{\mathscr{O}\left(x^3\right)}
\def\oxfour{\mathscr{O}\left(x^4\right)}
\def\oxfive{\mathscr{O}\left(x^5\right)}
\def\LL{\text{Lanczos-Lovelock}}

\def\ph{\phantom}

\begin{document}
\title{Signature of Non-uniform Area Quantization on Black Hole Echoes}
\author{Kabir Chakravarti}
\email{kabir.c@iitgn.ac.in}
\author{Rajes Ghosh}
\email{rajes.ghosh@iitgn.ac.in }
\affiliation{Indian Institute of Technology, Gandhinagar, Gujarat 382355, India.}
\author{Sudipta Sarkar}
\email{sudiptas@iitgn.ac.in}
\affiliation{Indian Institute of Technology, Gandhinagar, Gujarat 382355, India.}

\begin{abstract}
A classical black hole is characterized by a horizon that absorbs radiation of all frequencies incident on it. Perturbation of these black holes is well-understood via exponentially damped sinusoids known as quasi-normal modes. Any departure from such classical behavior near the horizon may induce significant modifications in the late time evolution of the perturbation leading to so-called gravitational wave echoes. This work considers the effect of black hole area-quantization on the formation of gravitational wave echoes. We investigate how the resulting echo waveform may depend on various model parameters. Our study opens up a new window to distinguish different models of area quantization using future gravitational wave observations and provides a novel probe to study the near horizon physics.

\end{abstract}
 
\maketitle 
\section{Introduction}
Gravitational wave (GW) astronomy is offering us with an intriguing opportunity to test the classical and quantum aspects of black holes \cite{LIGOScientific:2016aoc, LIGOScientific:2016lio}. In a recent work \cite{Agullo:2020hxe}, the authors have suggested that the GWs emitted from black hole inspirals may carry detectable imprints of the underlying quantum mechanical properties of the horizon. This is based upon Bekenstein's proposal \cite{Bekenstein:1974jk, Bekenstein:1995ju} of black hole area quantization: $A = \alpha\, l_{p}^{2}\, N$, where $l_p$ is the Planck length, $N$ is a positive integer, and $\alpha$ is a constant. Such discretization process, which could be a consequence of the Planck scale physics, may leave its signature on the emission \cite{Bekenstein:1995ju} as well as the absorption spectrum of black holes \cite{Foit:2016uxn, Cardoso:2019apo, Agullo:2020hxe, Datta:2020gem, Datta:2021row}.\\

\noindent
All these works crucially assume the validity of Bekenstein's entropy formula, i.e., the black hole entropy is proportional to the area of the event horizon. In general relativity (GR), this area law is motivated mainly from the Hawking's area theorem \cite{Hawking:1971tu}, which states that the area of a classical black hole can not decrease. However, once we venture beyond GR, the entropy of a black hole is no longer proportional to the horizon area, and can have sub-leading correction terms \cite{Wald:1993nt, Sarkar:2019xfd}. Also, if we interpret the black hole entropy as the entanglement entropy due to the entanglement of the modes of a quantum field, the area law can be obtained by tracing over the modes hidden by the black hole horizon \cite{Bombelli:1986rw}. Interestingly, if the quantum field is in a state different from the vacuum, the entropy receives sub-leading corrections  \cite{Das:2005ah, Das:2007mj, Sarkar:2007uz}. In fact, it is also proposed that, in general, only the entropy is quantized with an equally-spaced spectrum \cite{Kothawala:2008in}. If the entropy is not proportional to area, the horizon area is then quantized in a non-uniform manner.\\

\noindent
The upshots of such a non-uniform area quantization on the phasing of gravitational waveform from coalescing black hole inspirals is analysed in \cite{Chakravarti:2021jbv}. It has been shown that any correction to the area law may lead to detectable consequences in future GW observations. Moreover, such observations may as well put severe constraints on various parameters of the underlying model. This technique also provides a novel test for the area-entropy proportionality of black hole solutions in Einstein's theory of gravity.\\

\noindent
A natural question to ask is whether such area discretization may affect the post-merger ringdown phase. The ringdown spectrum for a classical black hole consists of the quasi-normal modes which are derived using the perfectly ingoing boundary condition at the horizon. However, due to its quantized area, a black hole can only absorb at certain characteristic frequencies. This would lead to a modification of the boundary condition at the horizon. Any such modification will affect the late time behaviour of the post-merger spectrum and gives rise to so-called GW echoes.\\

\noindent
GW echoes following the merger of compact objects have been investigated extensively in the last few years \cite{Cardoso:2016rao, Cardoso:2016oxy, Cardoso:2017cqb, Cardoso:2017njb, Mark:2017dnq, Correia:2018apm, Bueno:2017hyj}, and the possible presence of echoes in GW data are also being analysed    \cite{Abedi:2016hgu, Abedi:2017isz, Conklin:2017lwb, Westerweck:2017hus, Tsang:2018uie, Nielsen:2018lkf, Lo:2018sep}. Nevertheless, the modification of the horizon boundary condition which led to the generation of these echo signals were implemented in a rather ad hoc fashion. However, in \cite{Cardoso:2019apo}, it has been proposed that the quantization of black hole area may provide a concrete theoretical justification for the modified boundary condition at the horizon. It is modelled by adding a double-barrier potential near the horizon that mimics the selective absorption of quantum black holes. In that case, any future observation of GW echo in the late time signal from a binary black hole merger could confirm the hypothesis of area quantization.\\

\noindent
In this paper, we further develop the model proposed in \cite{Cardoso:2019apo} by introducing new perspectives. We demonstrate how the details of the area quantization can be captured by a careful choice of the double-barrier parameters that is placed to mimic the boundary condition on the horizon. It allows us to incorporate finer details of the area quantization in the echo spectrum, which was not manifest in the previous model \cite{Cardoso:2019apo}. As a consequence, our model breaks the universality of the echo-time, i.e., the time difference between two consecutive echoes, and make it dependent on the model of area quatization.\\

\noindent
Thus, our work compliments the results of \cite{Cardoso:2019apo}, by computing the possible effects of non-uniform area quatization on the gravitational wave echo spectrum similar to the inspiral case studied in \cite{Chakravarti:2021jbv}.  We use a scenario where the black hole entropy has sub-leading corrections in the form of power-law  \cite{Das:2005ah, Das:2007mj, Sarkar:2007uz}. Then, by assuming the entropy is quantized in equidistant steps, we can find the non-uniform quantization rules for the horizon area:

\begin{equation} \label{model}
A = \alpha\, l_{p}^{2}\, N\, \left ( 1 + C \, N^{\nu}\right)\, .
\end{equation}

\noindent
Here, $C$ is a constant present in the power-law correction term in the entropy. The other parameter $\nu$ is assumed to be negative in order to get back Bekenstein's area quantization law as $N \to \infty$. It is important to emphasise that in our model, the non-uniform area quantization need not be due to additional higher curvature terms; we continue using this model for GR black holes as well. Then, we analyse the effect of area discretization on black hole echo spectrum.\\

\noindent
Our work opens up a potential possibility to impose severe constraints on the choice of theoretically plausible models of area quantization using the future observation of black hole echo signals from different sources. Our study also provides another test for the area-entropy proportionality in parallel to the test presented in \cite{Chakravarti:2021jbv}.

\section{Quantum filter for non-uniform area quantization}

A classical black hole absorbs any radiation incident on it. In other words, the event horizon of a classical black hole has zero reflectivity ($R=0$) and unit transmissivity ($T=1$). However, the situation changes drastically when black hole's area is assumed to be quantized as prescribed in Eq.(\ref{model}). Then, it can only absorb at certain frequencies, characterized solely by the mass of a Schwarzschild black hole \cite{Agullo:2020hxe, Chakravarti:2021jbv}:

\begin{equation} \label{freq}
\omega_{N,n} = \frac{\alpha\, \kappa}{8 \pi}\, \left\{1 + C \left(1 + \nu\right) N^{\nu} \right\}\, n \, .
\end{equation}
\noindent
Here, $n$ is a positive integer, and $\kappa$ is the surface gravity at the event horizon. As a result, such quantum black holes will have frequency-dependent reflectivity $R(\omega)$.

\subsection{Gravitational Perturbation}
Although incorporating rotation in Eq.(\ref{freq}) is not a difficult job, we shall only focus on the non-rotating case. Then, our aim is to study perturbations on this black hole background as we modify the classical boundary condition at the horizon to model the absorption profile in terms of the characteristic frequencies given by Eq.(\ref{freq}). For this purpose, we consider a massless, quadrupole mode of the gravitational perturbation of the black hole. Then, the corresponding master equation for the perturbation $\Psi ( x, t)$ is

\begin{equation} \label{pert}
\left[ \partial_t^2 - \partial_x^2 + V_{\text{Sch}}\right] \Psi(x,t) = 0,
\end{equation}

\noindent
where the effective potential is denoted by, $V_{\text{Sch}} (r) = \left(6/r^2\right) (1-2M/r) (1-M/r)$. It is a well known fact that this potential has its maximum at $r = 3M$, which signifies the location of the light ring. As we shall see, this light ring will play a crucial role in the formation of black hole echo spectrum. Moreover, we have introduced the tortoise coordinate outside the horizon at $r=2M$ as $x(r) = r + 2M\, \text{log}\left(r/2M - 1\right)$.

\noindent
\\Now, using a Fourier transformation of the perturbation, $\Psi(x,t) := \int d\omega\,  \tilde{\Psi}(x,\omega)\, e^{-i \omega t}$, we can cast the master equation into its most useful form:

\begin{equation} \label{ft}
-\partial_x^2\, \tilde{\Psi}(x,\omega) = \left[ \omega^2 - V_{\text{Sch}}(x) \right] \tilde{\Psi}(x,\omega)\ .
\end{equation}

\noindent
It is interesting to notice the close resemblance of Eq.(\ref{ft}) with the time-independent Schrodinger equation. We want to study this equation with outgoing boundary condition near spatial infinity, i.e.,  $\tilde{\Psi} \propto e^{i \omega x}$ as $x \to \infty$. However, we should be careful in fixing the boundary condition near the horizon as the physics there will be modified due to the quantum effects sourced by the area quantization. It is reasonable to assume that these quantum modifications will only be important very close to the horizon, say upto a radius $r < r_{\epsilon} = 2M (1+\epsilon)$ for some positive values of $\epsilon << 1$.  For our purpose, we must put a reflecting boundary condition at $x = x_{\epsilon} := x(r_{\epsilon})$:

\begin{equation} \label{bc}
\tilde{\Psi}(x_{\epsilon},\omega) \propto e^{-i \omega\left(x-x_{\epsilon}\right)}+R(\omega)\, e^{i \omega\left(x-x_{\epsilon}\right)}\ ,
\end{equation}
\noindent
where $R(\omega)$ is the frequency-dependent reflectivity of the boundary at $r = r_{\epsilon}$. The classical result is obtained in the limit $R(\omega) = 0$.

\subsection{Modelling Quantum Filter}
\begin{figure} [h!]
\begin{center}
\includegraphics[width=6.5cm]{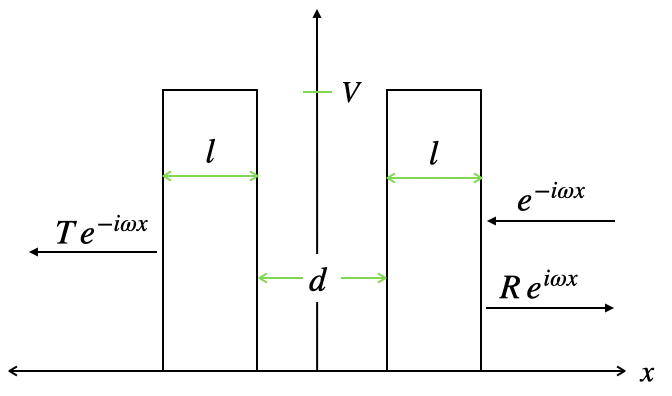}
\end{center}
\caption{Symmetric double-barrier potential with (rescaled) height $V$.}
\label{dwell}
\end{figure}

Quantization of black hole area demands that the absorption occurs only at the characteristics frequencies implying $R(\omega_{N, n}) = 0$, and otherwise $R(\omega) = 1$. Therefore, the absorption spectrum of a quantum black hole consists of a sharply peaked lines about its characteristic frequencies $\omega = \omega_{N,n}$ given by Eq.(\ref{freq}). As suggested in Ref.\cite{Cardoso:2019apo}, this behaviour can be modelled by a quantum mechanical double-barrier potential with a suitable choice of the barrier parameters, namely its height $V$, width $l$, and separation $d$ (See figure \ref{dwell}). However, in \cite{Cardoso:2019apo}, the area quantization is assumed to be uniform, whereas we are considering the effect of the non-uniformity.  \\

In order to mimic the black hole case, we must choose the double-barrier parameters so that the transmissivity $T(\omega)$ is very close to unity if $\omega = \omega_{N,n}$ given by Eq.(\ref{freq}). For any other frequency, $T(\omega)$ should be vanishingly small. For this purpose, it is instructive to calculate the transition amplitude $t(\omega)$ comparing the situation with the similar problem in quantum mechanics. 
\noindent
Using the result derived in \cite{Hauge:1987zz}, we may write (setting, $c=1$),

\begin{equation} \label{t}
t=\frac{e^{-i \omega(d+2 l)}}{e^{i \omega d}\mathcal{A}+e^{-i (\omega d-\delta)}\mathcal{B}}\, ,
\end{equation}

\noindent
with $\mathcal{A} =  \mathcal{M}^2\, \text{sinh}^2(\beta l)$, and $\mathcal{B} = \text{cosh}^2(\beta l) + \mathcal{K}^2\, \text{sinh}^2(\beta l)$. We have also defined, $\text{tan}(\delta) =  \mathcal{K}\, \text{sinh}(2 \beta l)\left[\text{cosh}^2 (\beta l) - \mathcal{K}^2\, \text{sinh}^2 (\beta l)\right]^{-1}$.\\

\noindent
Here, we are using a rescaled version of the potential in order to make the aforesaid comparison possible: $\left(2m/\hbar^2\right) U = V$, where $U$ is the actual potential that appears in the Schrodinger equation. Also, we have defined, $\beta^2 = V - \omega^2$. Since the rescaled energy of the particle, $\left(2m/\hbar^2\right) E = \omega^2$ is less than the height of the potential barrier $V$, the quantity $\beta$ is a positive real number. We have also used the notations: $2 \mathcal{M} = \beta/\omega + \omega/\beta$, and $2 \mathcal{K} = \beta/\omega - \omega/\beta$.\\
 
 \noindent
 Now, it is easy to calculate the transition probability which is defined by the equation, $T^2 = t\, \bar{t}$:

\begin{equation} \label{T}
T^{-2} = \mathcal{A}^2 + \mathcal{B}^2 +2\, \mathcal{A}\, \mathcal{B}\, \text{cos}\left(2\, \omega\, d - \delta \right)\ ,
\end{equation}

\noindent
Thus, the transition probability is an oscillatory function having two envelopes defined according to the maxima and minima of the sinusoidal part. The equations of these envelopes are as follows,

\begin{align} \label{env} 
\text{Upper envelope:}\, &\mid T \mid_u\, = \left(\mathcal{A} - \mathcal{B} \right)^{-1}= \, 1\, ;\nonumber \\
\text{Lower envelope:}\, &\mid T \mid_l\, = \left[ 2\, \mathcal{M}^2\, \text{sinh}^2(\beta l) + 1 \right]^{-1}\, .
\end{align}
 \noindent
 Therefore, this double-barrier model mimics the event horizon of the quantum Schwarzschild black hole if we can choose the barrier separation ($d$) so that $T(\omega_{N,n}) = T_u = 1$. This demand a condition on the barrier separation $d$ as $ 2\, \omega_{N,n}\, d \, -\, \delta = (2s+1)\, \pi$. Using this condition, we get
 
 \begin{align} \label{d}
 \omega_{N,n}= \frac{s\,\pi}{d} + \frac{\delta + \pi}{2d}\, .
 \end{align}
 
 \noindent
We should also check whether $T_l$ given in Eq.(\ref{env}) is vanishingly small so that $T(\omega) = T_l \approx 0$, when $\omega \neq \omega_{N,n}$. For a fixed value of the particle's energy, this can be assured by choosing larger and larger heights of the potential barrier, i.e., $V >> \omega^2$. In this limit, the phase of the sinusoidal part in transition amplitude becomes $\delta_r = r\, \pi$ for some integers $r$. Then, Eq (\ref{d}) gives

 \begin{align} \label{d1}
\omega_{N,n}\ = \left( s + \frac{r}{2}\right) \frac{\pi}{d} +  \omega_0\, , 
 \end{align}
 
 \noindent
where we define $\omega_0 = \pi /(2d)$. Next, we use Eq. (\ref{freq}) to obtain a direct relationship between the parameters of the double-barrier potential on the horizon and the quantization model of the area as,
 
 \begin{align} \label{d2}
 \frac{\alpha\, \kappa}{8 \pi}\, \left\{1 + C \left(1 + \nu\right) N^{\nu} \right\}\, n = \left( s + \frac{r}{2}\right) \frac{\pi}{d} +  \omega_0\, . 
 \end{align}
 
\noindent 
To proceed further, we note that our main purpose is to study the late time echo spectrum of the perturbation. For studying these echos, we shall consider the initial waveform to be a Gaussian that consists of all the modes $\omega_{N,n}$. Thus, our model makes better sense and simplifies if the barrier separation ($d$) does not carry any mode-index. In other words, the $n$-dependency must cancel out from both sides of the above equation. This task is achieved by first identifying the integers $n = r = s$, and then by making a constant shift in the frequency-scale, i.e., by redefining $\omega \to \omega + \omega_0$. \\

\noindent
Finally, Eq.(\ref{d2}) reads for non-uniform area quantization as, 

\begin{equation} \label{d3}
d = d_{\text{uniform}}\, \times \left[1 + C \left(1 + \nu\right) N^{\nu}\right]^{-1}\, ,
 \end{equation}                                                                                                                             
\noindent
where $d_{\text{uniform}} = 12\pi^2/(\alpha\, \kappa)$ is the corresponding quantity for uniform area quantization. This equation gives a one-to-one correspondence between the width of the double-barrier potential and the model of quantization labelled by the quantization parameters $(C, \nu)$. Note that the exact value of the quantity $\alpha$ depends on the details of the quantization. In his original proposal  \cite{Bekenstein:1974jk}, Bekenstein considered its value to be $8 \pi$, which is motivated by the transitions of a Schwarzschild black hole between different energy levels at discrete quasi-normal mode frequencies \cite{Hod, Maggiore}. However, other values of $\alpha$ are also studied in the literature \cite{Maggiore}.\\

\noindent
Interestingly, Eq.(\ref{d3}) may provide us an important tool to distinguish the effects of non-uniform quantization from that of uniform one in black hole echo. The uniform quantization gives a universal value for the quantity $\kappa\, d$, for all Schwarzschild black holes. This universality is lost once the quantization is non-uniform and its value depends on the black hole mass through $N$. Therefore, for a given value of $\alpha$, measuring the echo spectrum from multiple observations may provide us knowledge about the values of $d$. We can then check how different its value is from $d_{\text{uniform}}$. In fact, future echo observations may also be useful in putting stark bounds on model parameters $(C,\nu)$.\\

\noindent
One way to employ this idea is to plot the quantity $\kappa\, d$ for different values of $N$, which correspond to different black hole configurations. Then, the uniform qunatization is represented by a horizontal line parallel to the $N$-axis with an intercept of $\left(12\pi^2/\alpha \right)$ with the vertical axis. In contrast, any deviation from this horizontal line would indicate non-uniform quantization.\\

\noindent
If we demand the correction to the area law is only in the sub leading order,  Eq.(\ref{freq}) suggests that we should choose the values of $(C,\nu)$ so that $| C \left(1 + \nu \right) N^{\nu} |\, < 1$. Moreover, as discussed in \cite{Chakravarti:2021jbv}, for enriching the absorption spectrum, it is recommended to choose $C < 0$ rather than $C > 0$. All these considerations make the quantity $\left[1 + C \left(1 + \nu\right) N^{\nu}\right]$ to lie in the interval $(0,1)$. Interestingly from Eq.(\ref{d3}) we can infer, the more this quantity goes away from unity, the larger becomes the separation between $d$ and $d_{\text{uniform}}$ for the lower values of $N$ (correspond to black holes of a few solar mass). Thus, the sub-leading correction of area law forces the echo measurement for non-uniform quantizations to stand out vividly from the uniform one. However, as $N$ increases, the separation narrows down gradually to disappear at $N \to \infty$. 

\section{Black hole echoes}
Formation of black hole echoes crucially hinges upon two important ingredients: the event horizon and the light ring. Among them, the later one remains unaffected by the process of area quantization and it is located at $r = 3M$. The former one is modelled by the double-barrier potential to mimic the selective absorption spectrum of the black hole. Owing to this quantum filter, any perturbation of frequencies other than what are given by Eq.(\ref{freq}) will be reflected back from the rightmost barrier-wall of the potential, which signifies the extent of the quantum regime outside the black hole. It is important to note its dissimilarity from the reflection at a classical reflective surface that works as an amplitude-divider. In other words, any radiation irrespective of its frequency is partly reflected and partly transmitted through a classical barrier. In contrast, the horizon of an area-quantized black hole works as a frequency-filter. As a result, radiation of a particular frequency will be either absorbed or reflected completely, but not both simultaneously.\\

When the left-bound initial perturbation (taken in the form of a Gaussian waveform) reaches the light ring at $r=3M$, it excites the photon sphere modes. This, in turn, results in the initial ringdown signals.  After a certain time, these initial ringdown signals reflected back from the barrier and come back to the light ring, where it partially transmits through the potential maximum of $V_{\text{Sch}}$, see Eq.(\ref{ft}), and the remaining part is reflected back towards the horizon. As the process repeats itself, a series of black hole echoes is produced.\\

\subsection{Placing the Quantum Filter}
The duration between two consecutive echo signals, namely the echo-time is roughly given by the twice of the light travel time between the double-barrier boundary and the light ring. However, depending on details of the near-horizon physics, this separation can depend on the model of area quantization. In fact, we have two distinct ways of placing the double-barrier quantum filter near the horizon. As we shall see, these two perspectives will give rise to significantly distinct echo spectrum.\\

\begin{figure} [h!]
\includegraphics[width=6.5 cm]{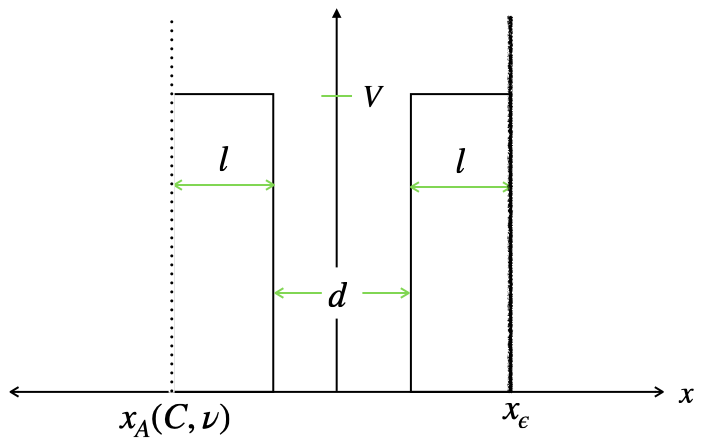}
\caption{Model 1: The dotted line represents the absorption surface that varies with $(C, \nu)$, and the thick line denotes the fixed reflection surface.}
\label{M1}
\end{figure}

The first model is closely related to what is depicted in \cite{Cardoso:2019apo}, where the rightmost barrier-wall of the potential is aligned with the surface of quantum extent located at $x=x_\epsilon$. In this model, we assume the location of $x_\epsilon$ that acts as a reflecting surface is model-independent, and can be fixed universally for all Schwarzschild black holes of mass $M$. However, due to the model-dependency of the barrier separation $d$ given by Eq. (\ref{d3}), the location of the inner barrier-wall may vary, see Fig. [\ref{M1}]. We can think of the inner barrier-wall as the location of the modified absorption surface $x_A(C, \nu)$ due to the quantum effects near the horizon. Since the distance between the reflecting surface $x_\epsilon$ and the location of the light ring does not vary with different choices of $(C, \nu)$, the echo-time is independent of the nature of area quantization. In other words, non-uniform area quantization leads to the same echo-time as that of uniform quantization.\\

\begin{figure} [h!]
\includegraphics[width=6.5 cm]{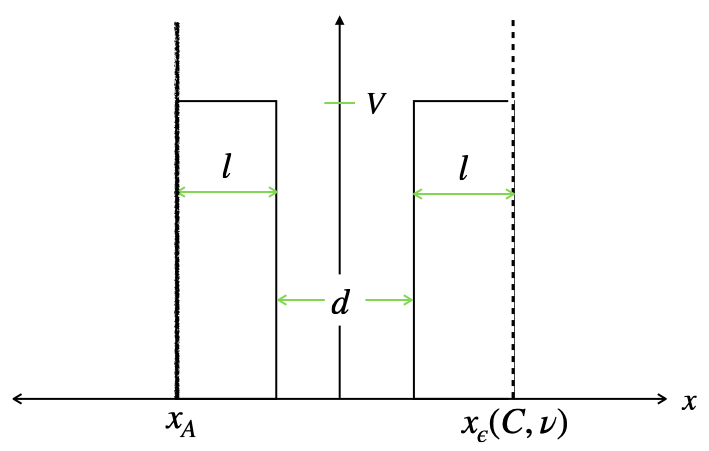}
\caption{Model 2: The dotted line represents the reflection surface that varies with $(C, \nu)$, and the thick line denotes the fixed absorption surface.}
\label{M2}
\end{figure}

\noindent
The universality of the echo-time can be lifted by introducing an alternative perspective. In this model, we consider the location of the absorption surface at $x_A$ to be fixed and model-independent. However, the location of the reflecting surface $x_\epsilon$ is now varying due to the change of the barrier separation $d$ as given by Eq. (\ref{d3}), for different choices of model parameters $(C, \nu)$, see Fig. [\ref{M2}] . As a result, the separation between $x_\epsilon$ and the light ring becomes model-dependent and so is the echo-time.

\noindent
However, in any case, we need to be careful so that the values of $x_A$ and $x_\epsilon$ is indeed close to the horizon and inside the quantum regime.

\subsection{Cauchy Evolution Equation}
In both the models, to find the echos, we need to express the time-domain master equation for the perturbation in double null coordinates defined by $u := t-x,\, \text{and}\, \, v := t+x$, in the units of $c=1$. In this coordinates, Eq. (\ref{pert}) takes the form

\begin{equation} \label{null}
\left[ 4\, \partial_u\, \partial_v + V \right] \Psi(u,v)= 0\ .
\end{equation}

\noindent
Here, the potential $V$ consists of two parts, the classical Schwarzschild potential $V_{\text{Sch}}$, and the quantum filter $V_{\text{Barrier}}$ near the horizon. The second part is required to incorporate the quantum boundary condition at the boundary at $x = x_{\epsilon}$. Thus, the perturbation evolves under the influence of the combined potential $V = V_{\text{Sch}} + V_{\text{Barrier}}$. To solve Eq. (\ref{null}) numerically, we discretize the $(u,v)$-plane in the form of a square grid of length $h << 1$. Then, the evolution equation becomes,

\begin{align} \label{dis}
&\Psi(u+h,\, v+h)=  \Psi(u+h,\, v) + \Psi(u,\, v+h) - \Psi(u,v)& \nonumber \\ 
&- \frac{h^2}{8} \left[V(u+h,v)\Psi(u+h,v) + V(u,v+h)\Psi(u,v+h)\right].
\end{align}

\noindent
Using this equation, we can now study the evolution of an initial Gaussian waveform in order to find the black hole echo spectrum.\\

\subsection{The Echo Spectrum}

We begin with an analysis of the echo-time $\Delta t_{echo}$ for both the models. As discussed before, the echo-time is roughly twice the light travel time between the light ring and the rightmost barrier of the quantum filter. In the leading order, it is given by 

\begin{equation} \label{echot}
M^{-1} \Delta t_{echo} \sim 2 \left[ 1 - 2\, \text{ln} 2 - 2\epsilon + 2\, \text{ln}(\epsilon^{-1})\right]\, .
\end{equation}

\noindent
Note that the quantity $M^{-1} \Delta t_{echo}$ has no explicit dependence on mass $M$; the only dependence on mass may come implicitly through the parameter $\epsilon$ in the context of model-2.\\

\noindent
In model-2, the absorption surface is fixed at a distance $r_A = 2M (1+a)$, where $a  < \epsilon < < 1$. Then, the extent of the quantum region is governed by the following equation: 

\begin{equation} \label{ea}
\epsilon(C,\, \nu) = a\, \text{Exp} \left[\frac{24 \pi^2}{\alpha \left(1 + C N^{\nu} (1 + \nu)\right)} \right]\, .
\end{equation}

\noindent
Thus, for a given mass $M$, the quantum-extent outside the horizon ($\epsilon$) can vary with different choice of quantization parameters $(C,\, \nu)$. Also, for a fixed choice of the parameters  $(C \neq 0,\, \nu)$, the extent $\epsilon$ will be different for different masses via $N$.\\

\noindent
In contrast, for model-1, $\epsilon$ is a constant parameter independent of $(M, C, \nu)$, and fixed only by the quantum extent outside the event horizon. We also point out that in both the models, Eq.(\ref{echot}) gives the same value of $M^{-1} \Delta t_{echo}$ for uniform quantization ($C =0$).\\

\begin{figure} [h!]
\includegraphics[width=8cm]{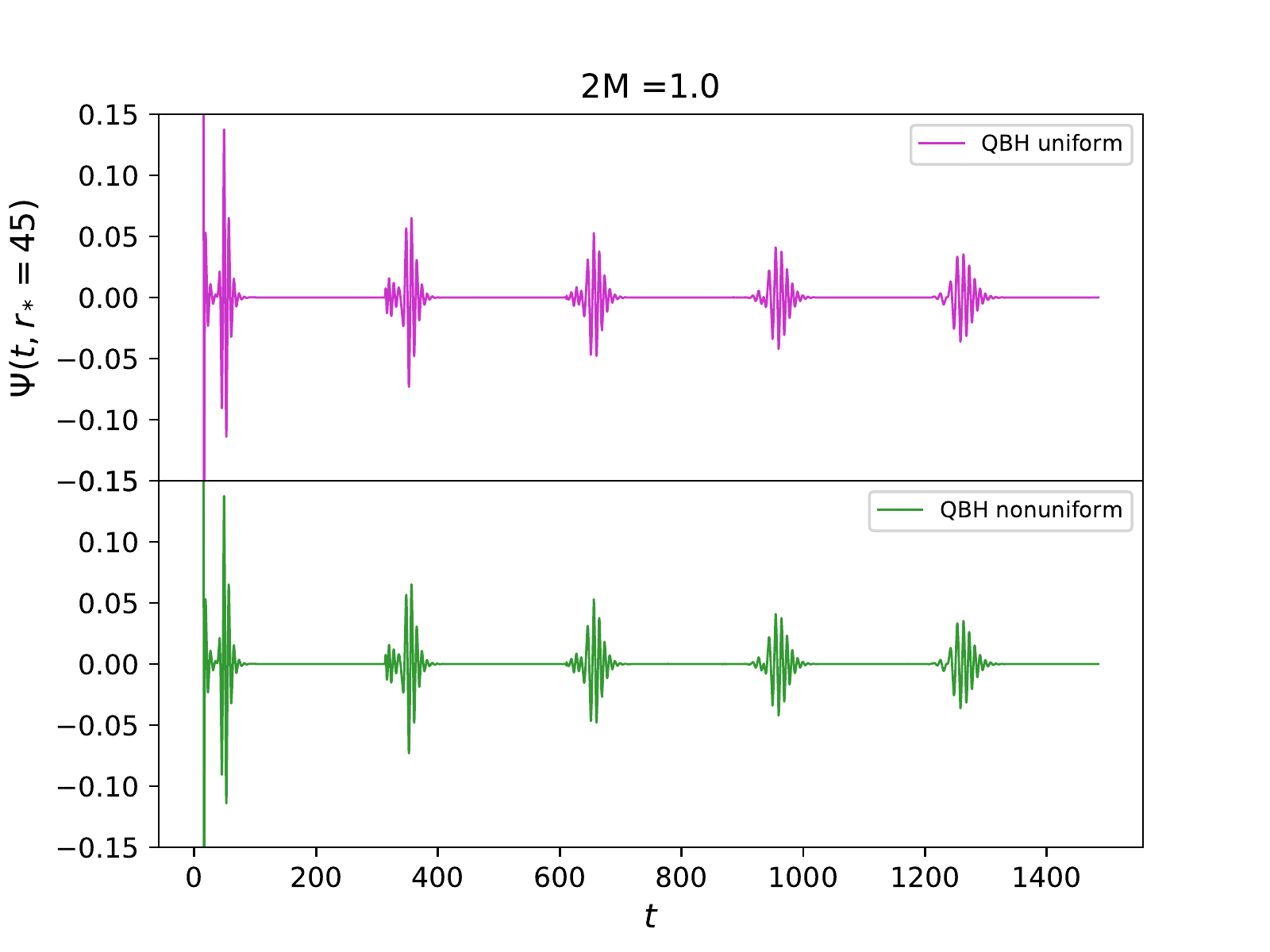}
\caption{Gravitational wave echo spectrum for model-1. The parameters for the non-uniform quantization are $C = -3.6$, and $\nu = -1/90$. We have chosen $\epsilon = 10^{-59} $. }
\label{P1}
\end{figure}

\noindent
Now, let us consider the echo spectrum for both the models separately. Fig.[\ref{P1}] shows the echo spectrum  corresponding to model-1, for both uniform and non-uniform area quantizations. As it is expected, the echo wavefronts are almost identical revealing no information about the model parameters $C$ and $\nu$. This is because the distance between the reflecting surface $x_\epsilon$ and the location of the light ring does not vary with $(C, \nu)$. As a consequence, the echo-time $\Delta t_{echo}$ is unaltered for different models and depends only on the mass $M$ of the black hole. In fact, in the leading order, the quantity $M^{-1} \Delta t_{echo}$ does not depend on the mass. \\

\begin{figure} [h!]
\includegraphics[width=8cm]{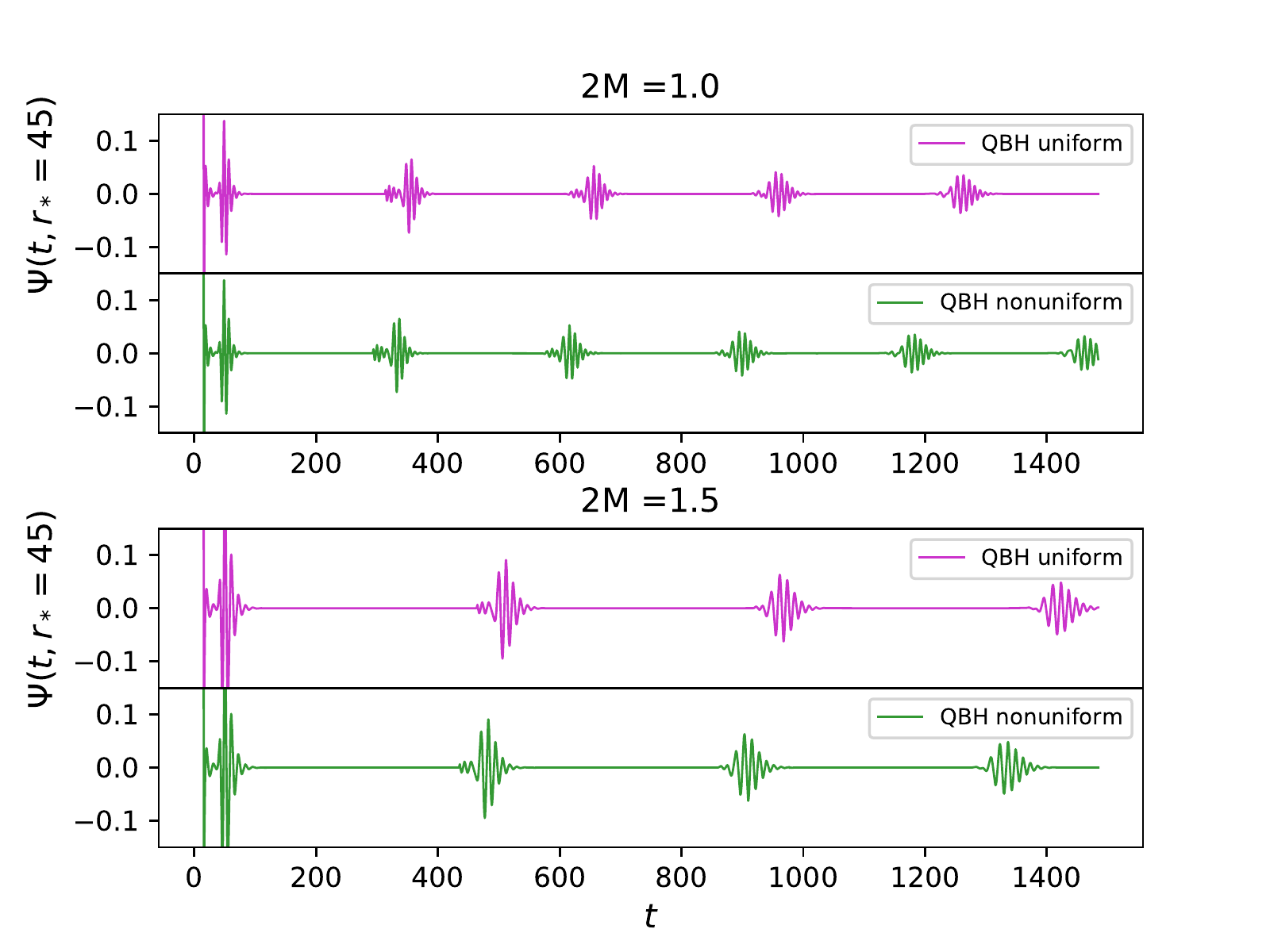}
\caption{Gravitational wave echo spectrum for model-2. The parameters for the non-uniform quantization are $C = -3.6$, and $\nu = -1/90$. We have chosen $a = 10^{-70} $. }
\label{P2}
\end{figure}

The situation for model-2 is depicted in Fig.[\ref{P2}]. This is the case when the location of the absorption surface at $x_A$ is fixed and model-independent, and the reflecting surface $x_\epsilon$ is varying due to different choices of model parameters $(C, \nu)$. As a result, the echo-time, the time separation between two consecutive echoes depends on the nature of quantization. In fact, this immediately suggests a possible test to distinguish these two models.\\

\noindent
In the context of model-2, if the gravitational echo is observed by future detectors for more than one GW sources, a significant variation of the quantity $ M^{-1} \, \Delta t_{\textrm{echo}}$ will be a strong evidence for non-uniform area quantization. This is quite explicit in an order-of-magnitude estimation of the echo time for model-2, $M^{-1} \Delta t_{echo} \sim  4\, \text{ln} \left( \epsilon_{\text{max}} /  \epsilon \right).$ Here, the quantity $\epsilon_{\text{max}} = \sqrt{e} / 2$ can be fixed universally. Interestingly, the positivity of echo-time gives an upper bound on $\epsilon$ for a given choice of $(C, \nu)$: $\epsilon < \epsilon_{\text{max}}$. This bound can be translated to a upper bound on the parameter $a$ that signifies the location of the absorption surface. 

 \begin{equation} \label{em}
a_{\text{max}} = \frac{\sqrt{e}}{2}\, \text{Exp} \left[-\frac{24 \pi^2}{\alpha \left(1 + C N^{\nu} (1 + \nu)\right)} \right]\ .
 \end{equation}

There is no such bound on $\epsilon$ for model-1, since the echo-time is always positive.\\

\begin{figure}[h!]
\includegraphics[width=7cm]{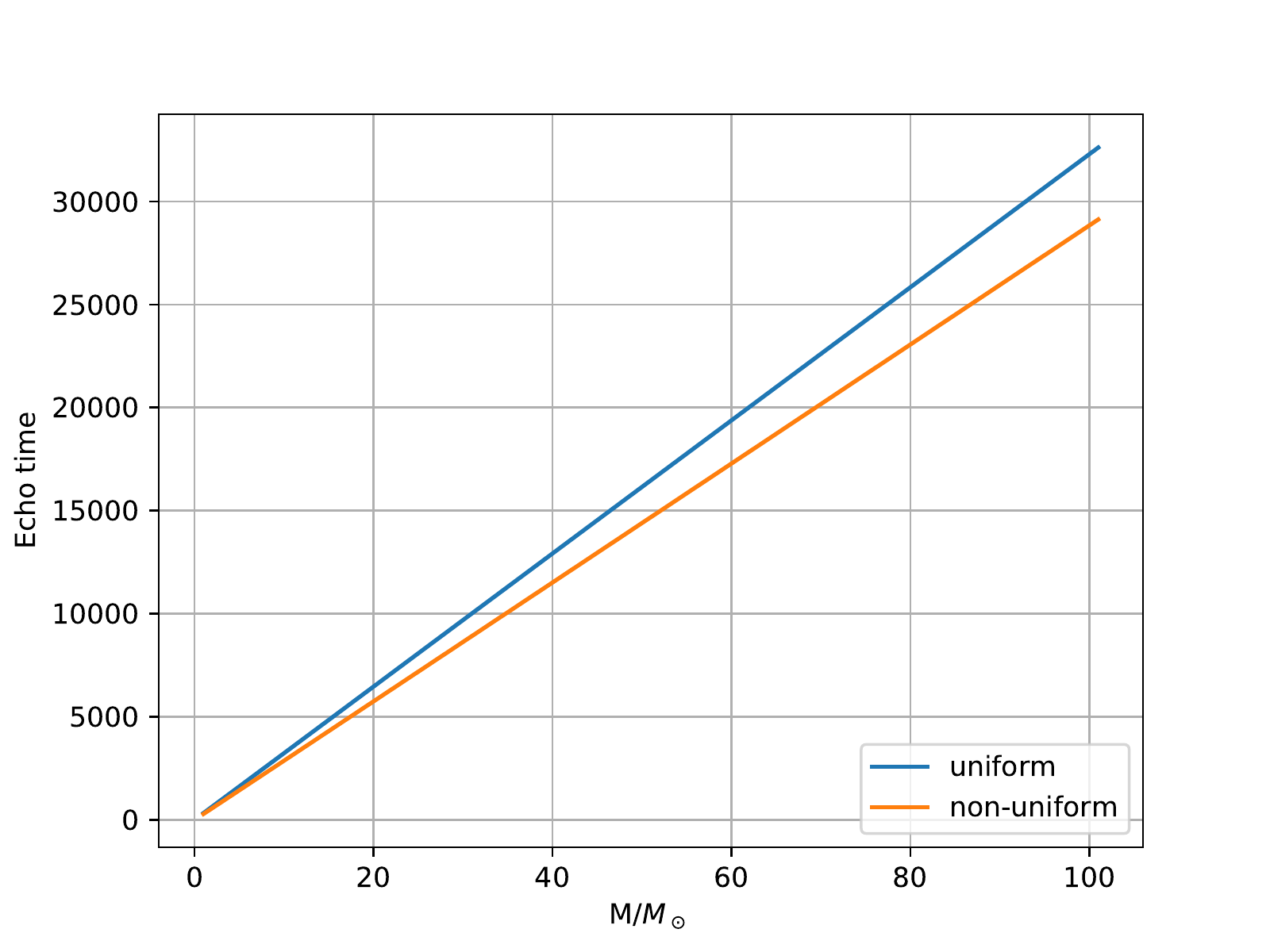}
\caption{Variation of echo-time (in units of solar mass $M_{\odot}$) with mass of the black hole for model-2. The parameters for the non-uniform quantization are $C = -3.6$, and $\nu = -1/90$.}
\label{t1}
\end{figure}

\noindent
Fig. [\ref{t1}] shows the variation of the echo-time with black hole mass for model-2. There is a noticeable difference between uniform and non-uniform area quantizations for higher values of the black hole mass. \\ 

\noindent
In order to distinguish between these two models depending on the placement of quantum filter near the horizon, and also between the uniform and non-uniform area quantization, we may proceed as follows. First, from the gravitational echo observations by future detectors, we may investigate the variation of the quantity $ M^{-1} \, \Delta t_{\textrm{echo}}$ for multiple GW sources of different masses. If it varies significantly for different sources, then model-2 is preferred over model-1. Subsequently, we try to fit the observed echo-time data with theoretically expected values in model-2 for different choices of $(C, \nu)$. If the best fit model parameter significantly differ from $C=0$, it would suggest that the black hole area quantization is non-uniform.

\section{Conclusion}

In this paper, we have explored the effects of both uniform and non-uniform area quantization on GW echoes from a perturbed black hole. As a result of area discretization, black holes deviate from their classical behavior and absorb selectively at certain characteristic frequencies. Motivated by the method presented in \cite{Cardoso:2019apo}, we model this selective absorption using a double-barrier potential placed near the horizon. We show, depending on the near-horizon physics, there are two distinct ways to place the potential barrier, which lead to drastically different consequences on the echo spectrum. The echo profile, in particular the echo-time is sensitive to the quatization parameters $(C, \nu)$, only for one of these models, namely model-2. In that case, the echo-time $\Delta t_{echo}$ scales non linearly with the mass and potentially offers an observational test to distinguish these two models. In fact, since $\Delta t_{echo}$ does not depend on the parameters $(C, \nu)$ for model-1, a parameter estimation for $(C, \nu)$ should return only $C = 0$. On the other hand, model-2 may allow non-zero values of $C$. \\

\noindent
Note, we have only considered power-law corrections to the area-entropy proportionality, motivated by the works \cite{Das:2005ah, Das:2007mj, Sarkar:2007uz}. Interestingly, a logarithmic correction to the entropy will not have any detectable effects. This is similar to the inspiral case, as pointed out in \cite{Chakravarti:2021jbv}.\\

\noindent
It will be interesting if we can generalize our analysis beyond spherical symmetry and for more general black hole spacetimes. Also, extension for black holes of higher curvature gravity theories may also be useful.

\section*{Acknowledgement}

K.C. acknowledges the use of the cluster Noether at IIT Gandhinagar.
The research of R.G. is supported by the Prime Minister Research Fellowship (PMRF-192002-120), Government of India. The research of S.S. is supported by the
Department of Science and Technology, Government of
India under the SERB CRG Grant (CRG/2020/004562).

\end{document}